\documentclass[twocolumn,prl,superscriptaddress,preprintnumbers,amsmath,amssymb]{revtex4}
\usepackage{graphicx}
\usepackage{color}
\usepackage{amssymb}

\def\gtrless{\raise2.5pt\hbox{$>$}\llap{\lower2.5pt\hbox{$<$}}}
\def\gtrapprox{\raise2.5pt\hbox{$>$}\llap{\lower2.5pt\hbox{$\approx$}}}

\newcommand{\bsq}[1]{\begin{subequations}\label{#1}}
\newcommand{\esq}{\end{subequations}}
\newcommand{\beq}[1]{\begin{equation}\label{#1}}
\newcommand{\eeq}{\end{equation}}
\newcommand{\beqa}[1]{\begin{eqnarray}\label{#1}}
\newcommand{\eeqa}{\end{eqnarray}}

\newcommand{\rem}[1]{}

\newcommand{\rb}{{\bf r}}
\newcommand{\qb}{{\bf q}}

\newcommand{\ub}{{\bf u}}

\renewcommand{\rho}{\varrho}
\renewcommand{\epsilon}{\varepsilon}

\begin{document}

\title{Strain pattern in supercooled liquids}

\author{Bernd Illing}
\affiliation{University of Konstanz, D-78457 Konstanz, Germany}
\author{Sebastian Fritschi	}
\affiliation{University of Konstanz, D-78457 Konstanz, Germany}
\author{David Hajnal}
\affiliation{Johannes Gutenberg-University Mainz, D-55099 Mainz, Germany\\
Present Adress: BASF SE, Carl-Bosch-Strasse 38, D-67056 Ludwigshafen, Germany}
\author{Christian Klix}
\author{Peter Keim}
\affiliation{University of Konstanz, D-78457 Konstanz, Germany}
\author{Matthias Fuchs}
\affiliation{University of Konstanz, D-78457 Konstanz, Germany}

\date{\today}

\begin{abstract}
Investigations of strain correlations at the glass transition reveal unexpected phenomena. The shear strain fluctuations show an Eshelby-strain pattern ($\,\sim \cos{(4\theta)}/r^2\,$), characteristic for elastic response, even in liquids at long times \cite{Lemaitre2013}. We address this using a mode-coupling theory for the strain fluctuations in supercooled liquids and data from both,  video microscopy of a two-dimensional colloidal glass former and simulations of Brownian hard disks. We show that long-ranged and long-lived strain-signatures follow a scaling law valid close to the glass transition.  For large enough viscosities,  the Eshelby-strain pattern is visible  even on time scales longer than the structural relaxation time $\tau$ and after the shear modulus has relaxed to zero.

\end{abstract}


\maketitle

Glasses behave like isotropic elastic solids under external loads.  Their strain fields are long-ranged as captured in elasticity theory. Considering appropriate boundary conditions, Eshelby obtained the elastic strain field surrounding an isolated local deformation  \cite{Eshelby1957}, which is also the basis for modern theories of plasticity in disordered solids \cite{Barrat2011}:  Plastic deformation proceeds via localized irreversible rearrangements coupled by elastic strain fields.

While the relevance of strain in glass at low temperatures originates in the breaking of translational symmetry underlying solidification \cite{Martin1972}, the proper understanding of the evolution of strains at the crossover from metastable glass to supercooled liquid remains an open topic \cite{Flenner2015b}. The concept of plastic events, which are elastically coupled,  has emerged as one candidate rooted in the theory of solids which aims to capture the glass transition from the low temperature side. It suggests that `supercooled liquids are solids that flow' \cite{Dyre2006} and focuses on strain fluctuations and their correlations to probe plastic flow.

Lema\^{i}tre and colleagues  found evidence for this concept in molecular dynamics  simulations of two-dimensional Lennard-Jones mixtures: They observed persistent long-ranged strain fluctuations in supercooled liquid states \cite{Lemaitre2013}. Because the time over which strains were accumulated exceeds the structural relaxation time $\tau$ of the liquid, observable elastic stresses have decayed. The observation of spatial dependences exhibiting a far-field Eshelby-strain pattern thus can not be a simple consequence of elasticity. It suggests that particle rearrangements in a fluid interact over large distances via strains likely in an underlying elastic structure,  the `inherent states' characterizing the potential energy landscape \cite{Lemaitre2014}.\rem{ It also questions the concept of dynamical length scales  \cite{Karmakar2016}.}

Strain patterns have been observed experimentally, but not yet in quiescent supercooled liquids.
An anisotropic decay of strain was found in a 3D colloidal hard sphere glass under steady shear \cite{ChikkadiEPL2012}, in 2D simulations \cite{Varnik2014}, and in granular matter \cite{Denisov2016}. Eshelby-patterns are reported in 2D flowing emulsions \cite{Desmond2015} and in a 3D colloidal hard sphere glass, where they appear under shear  and thermally induced in quiescent state \cite{Jensen2014}. They are also present in 2D soft hexagonal crystals with dipolar interaction \cite{Franzrahe2008a}. Simulations revealed Eshelby-patterns in a 2D flowing foam \cite{Chikkadi2015} and in  a glass-forming mixture under shear \cite{Nicolas2014}. For a monodisperse fluid system in 2D of particles with screened Coulomb interaction (this system crystalizes at low temperatures), these  patterns were observed even at temperatures above the hexatic phase in the short time regime due to the  high frequency shear modulus \cite{Wu2015}.

In this letter, we present the first experimental evidence for Eshelby-like strain patterns in quiescent supercooled liquids and provide a theoretical description rooted in theories of liquid dynamics. This establishes the dissipative transport mechanism leading to long-ranged strain fluctuations and identifies the spatial and temporal window where they can be observed in supercooled states. We determine the viscous parameters quantifying the strains in liquids and derive a scaling law  for the crossover from glassy to liquid dynamics.

 Monolayers of binary mixtures of dipolar colloids have emerged as model system for the study of the glass transition by video microscopy \cite{Koenig2005}. The dipolar interaction between colloids can be tuned by an external magnetic field and the interaction parameter $\Gamma$ (magnetic over thermal energy) is precisely known; it may be considered a dimensionless inverse temperature. Strain fields can be determined from the particle trajectories in crystalline \cite{Keim2004} and amorphous  \cite{Klix2012} solids, and the jump of the shear modulus $\mu$ at vitrification could be measured \cite{Klix2015}. Solutions of the mode coupling theory (MCT) provide theoretical results for the glass transition  \cite{Hajnal2011}.

The dynamics of the system shall be captured in the  collective mean-squared  strains  at two different locations  $\rb_{1,2}$\rem{ accumulated over time $t$},
${C}_{\alpha\beta\gamma\delta}(\rb_{1},\rb_{2},t) =   \langle
\Delta\epsilon_{\alpha\beta}(\rb_1,t) \; \Delta\epsilon_{\gamma\delta}(\rb_2,t)   \rangle$.
Here, $\epsilon_{\alpha\beta}$ is the familiar (linearized) strain-tensor with spatial indices $\alpha,\beta \in \{1,2\}$ in two dimensions. It is obtained from the differences of particle positions accumulated over the time span $t$ \cite{Goldhirsch2002}.
Equilibrium averaging is done with the Gibbs-Boltzmann distribution \cite{Hansen}, and  time-translational invariance of the system is assumed and experimentally achieved by careful equilibration for several days up to weeks in between the measurements at different temperatures ($\Gamma$). In an homogeneous system, the correlations depend on the distance $\rb=\rb_2\!-\!\rb_1$ only.
In an isotropic system, the fourth rank tensor can be reduced to two independent functions related to compressional and shear deformations.  Following Ref.~\cite{Lemaitre2013}, we focus on the transverse  element $C_{xy}(\rb,t) = {C}_{xyxy}(\rb,t)$; more details can be found in the supplemental material (SI) \cite{SI}.

\begin{figure}[tb]
\includegraphics[width=\linewidth]{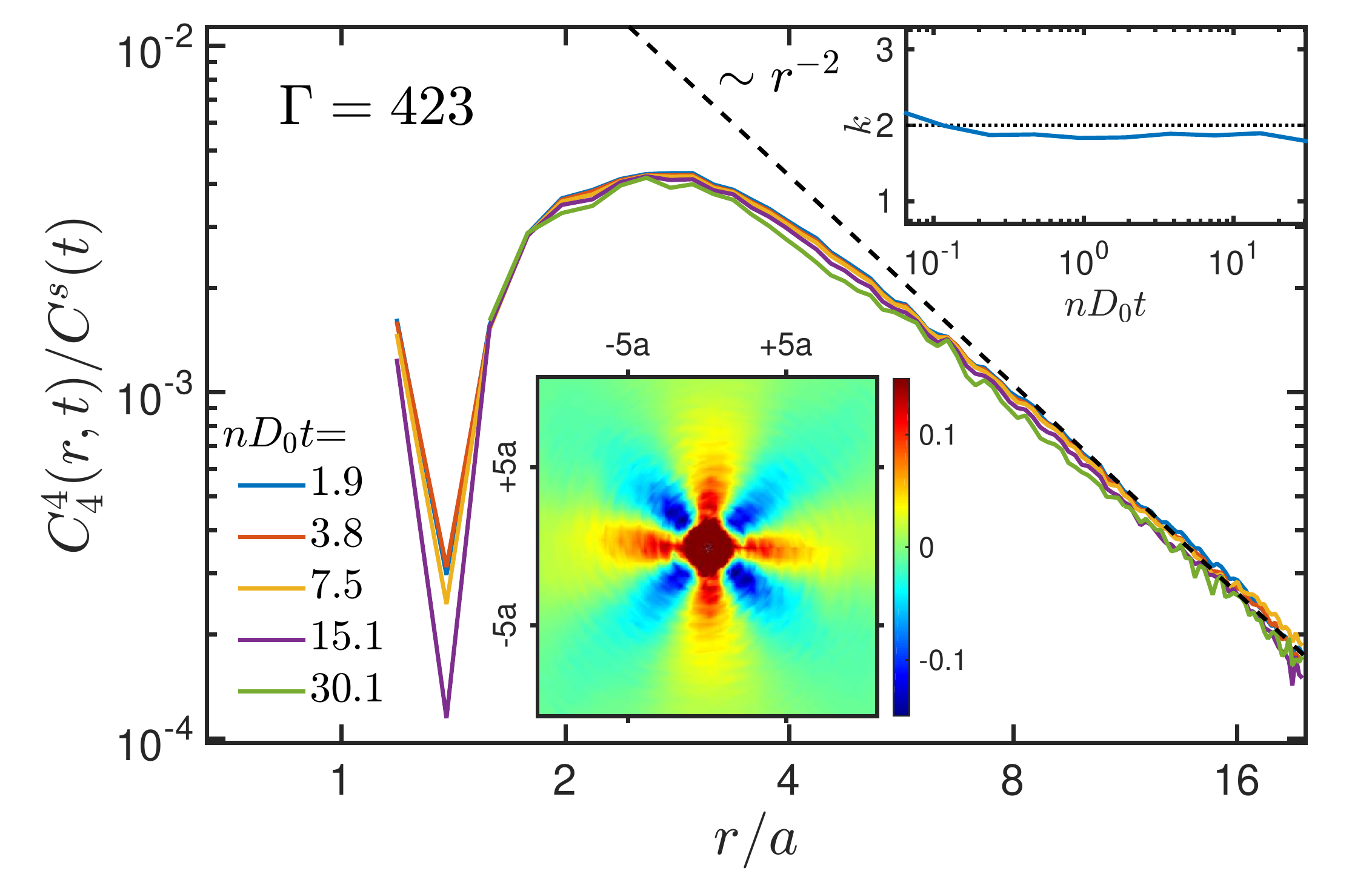}
\includegraphics[width=\linewidth]{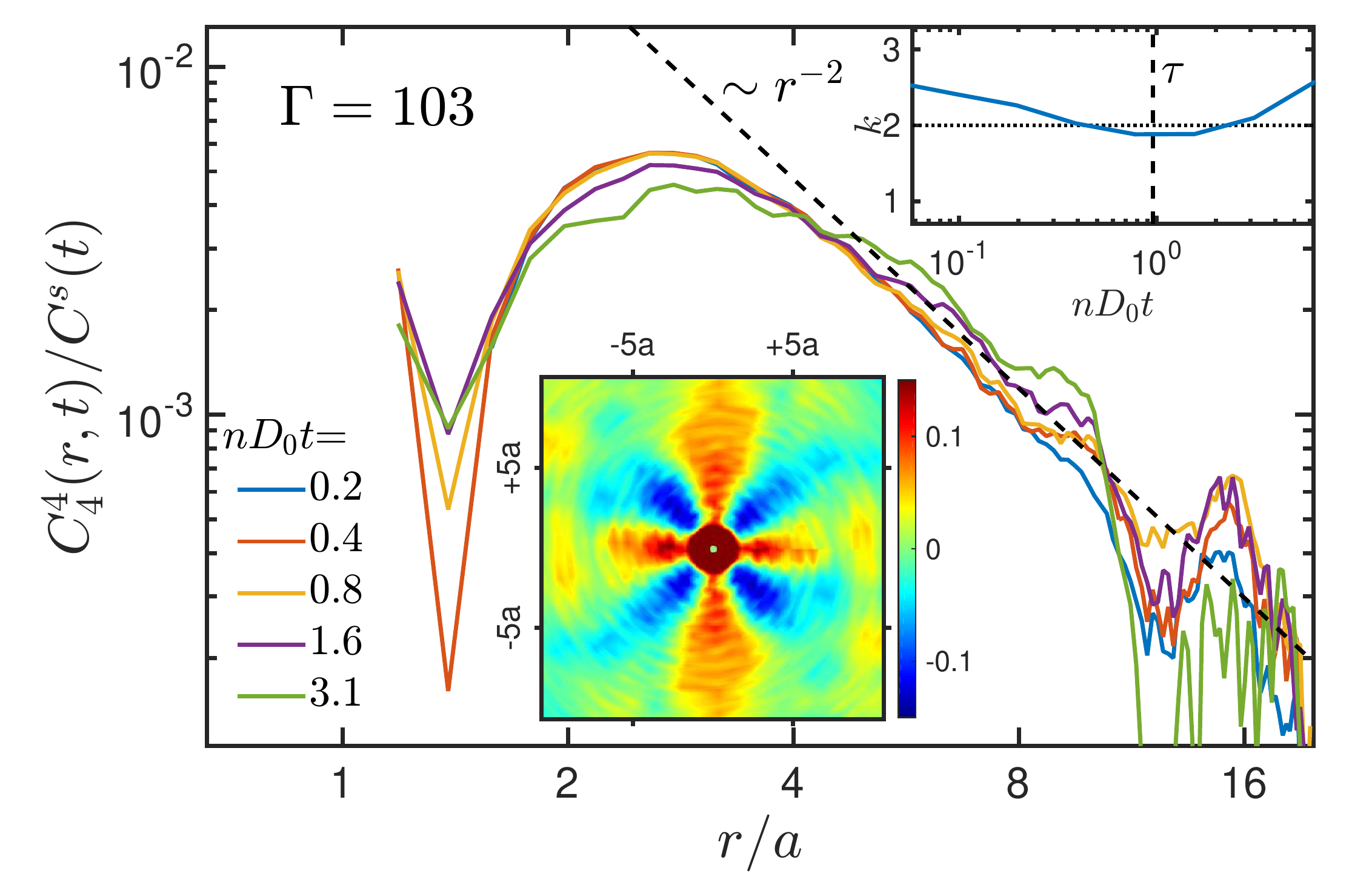}
\caption{Experimental rescaled strain correlation data for a glass (upper panel; $\Gamma=423$) and a fluid (lower panel; $\Gamma=103$) state at different times (see legends). The  spherical harmonic strain correlation functions $C_4^4({ r},t)/C^s(t)$ are rescaled to overlap in the far-field power law decay.  Main panels show the $1/r^k$-power law decay (dashed black), with exponent $k = 2$ varying little with time (upper insets). The contourplots (lower insets) of the long-time limit of $C_{xy}({\bf r},t)/C_{xy}({\bf r}=0,t)$ illustrate the corresponding $\cos{(4\theta)}$-symmetry; The Eshelby-patterns are shown at $nD_0t=30.1$ (glass) and $nD_0t=3.1 > nD_0\tau$ (fluid), respectively.}
\label{fig1}
\end{figure}

Figure~\ref{fig1} shows the measured transversal mean-squared strain fluctuations for a glass and a fluid state. They were obtained using video-microscopy on a binary colloidal monolayer \cite{Klix2015} following the analysis introduced in Ref.~\cite{Goldhirsch2002}. A brief introduction of the experimental setup can be found in the SI \cite{SI} while an elaborate discussion is given in \cite{Ebert2009}. Time is given in reduced units with $D_0n$ the rate for (unhindered) diffusion of a particle over the average particle separation $a=1/\sqrt{n}$, where $n$ is the particle density; $D_0$ is the dilute diffusion coefficient.
One notices  the angular dependence of the far-field strain, which Eshelby obtained for the elastic distortion around a localized disturbance at the origin \cite{Eshelby1957,Picard2004}.
\begin{equation}\label{e2}
   C_{xy}(\rb,t)  \to \cos{(4\theta)}\; \frac{C^s(t)}{4\pi n r^2} \quad  \mbox{for}\; r\gg a\;.
\end{equation} Four lobes of maximal intensity alternate with four lobes of minimal intensity.
 The appropriate spherical harmonics projection,
 $ C_4^4(r,t) =\frac{1}{\pi}\int_{0}^{2\pi}
  d\theta \cos(4\theta) C_{xy}(\mathbf{r},t)$,  decays slowly  at large separations $r\gg a$, with a power-law  $C_4^4 \propto r^{-k}$ of exponent $k=2$ (see dashed line in Fig.~\ref{fig1}).
This is the classical result from continuum mechanics and linear response theory (viz.~the fluctuation dissipation theorem) for the strain fluctuations in an isotropic solid. We find that it holds for times beyond the short time local dynamics (viz.~$n D_0 t \gtrapprox 1$) and  for distances $r$ larger than the average particle separation $a$. The algebraic decay of hexadecupolar symmetry follows from the fundamental equation of elastostatics, which predicts for the amplitude of the algebraic decay  $C^s(t\to\infty)=2k_BT n\left({1}/{\mu}-{1}/{\mu^\|}\right)$. Here, the elasticity seen  in (volume-preserving) strain-deformations is the hallmark of a solid and results from a finite shear modulus $\mu$. The longitudinal modulus $\mu^\|$, which would be present in a fluid also, gives a (small) correction in $C_{xy}$. The observation of a finite shear rigidity  is consistent with the interpretation  that the colloidal layer is in a solid state at low temperatures, viz.~$\Gamma> \Gamma_g \approx 200$, where $\Gamma_g$ is the inverse dimensionless glass transition
temperature obtained from the discontinuity in the elastic moduli \cite{Klix2015}.

The measured spatial correlations of the strains persist in fluid states at $\Gamma = 103 < \Gamma_g$.  At this temperature, approximately twice higher than the glass transition temperature, the mean-squared strains exhibit spatial correlations reminiscent of solids even for times far larger than the structural relaxation time $\tau$. The upper inset in Fig.~\ref{fig1}b compares the times $t$ where Eq.~\eqref{e2} holds with $\tau$. It is estimated from the decay of density correlations with wavelength of the average particle separation; the precise definition from the normalized collective intermediate scattering function $\Phi_{qa\approx 2\pi}(t)$ is presented in the SI \cite{SI}. The relaxation time $\tau$ also characterizes the decay to zero of the shear stress auto-correlation function, which indicates
that the fluid can not sustain elastic shear stresses for such long times \footnote{While noise on the measured interaction forces prevented us from obtaining the shear stress auto-correlation functions in experiment, the BD simulations provide $\tau$ from density and stress correlations and show the validity of Eq.~\eqref{e2} up to $nD_0t\approx500 >40\tau$ (for $\phi=0.78$) \cite{SI}.}. Thus we experimentally recover the intriguing observation by Lema\^{i}tre and colleagues that solid-like Eshelby strain fields survive in supercooled fluids even though density and stress correlations are fluid like.

In order to understand the spatial strain correlations at the glass transition, we turn to microscopic quantities which provide insights into fluid and solid states. The transversal collective mean-squared displacement (TCMSD) function
 ${ C }^{\perp}(q,t) =   \langle \Delta \ub_\qb^{\perp \ast}(t)\,\Delta \ub_\qb^\perp (t)  \rangle$
can be obtained from the particles' displacements accumulated in the time interval $t$ \cite{Klix2012}. The plane wave decomposition given by the spatial Fourier-transformation is possible in homogeneous systems and allows to focus on shear fluctuations perpendicular to the wavevector $\qb$.
The TCMSD may be considered a collective and spatially resolved generalization of the single particle mean squared displacement (MSD) familiar from liquid dynamics \cite{Hansen}.

 The transverse strain correlation function is given by  inverse Fourier-transformation: $C_{xy}(\rb,t) =FT^{-1}\left[ (  \frac{ q_x^2+q_y^2}{4}-\frac{q_x^2q_y^2}{q^2} ) C^{\perp}(q,t)\right] (\rb)$,
 where the $\qb$-dependent factors arise as strains are the symmetrized gradients of displacement fields in linear order \cite{Salencon}. For simplicity of presentation, we assume an incompressible system from now on, neglecting a longitudinal CMSD $C^{\|}(q,t)$, and relegate the complete theoretical analysis of compressible systems to the SI \cite{SI}.

\begin{figure}[tb]
\includegraphics[width=\linewidth]{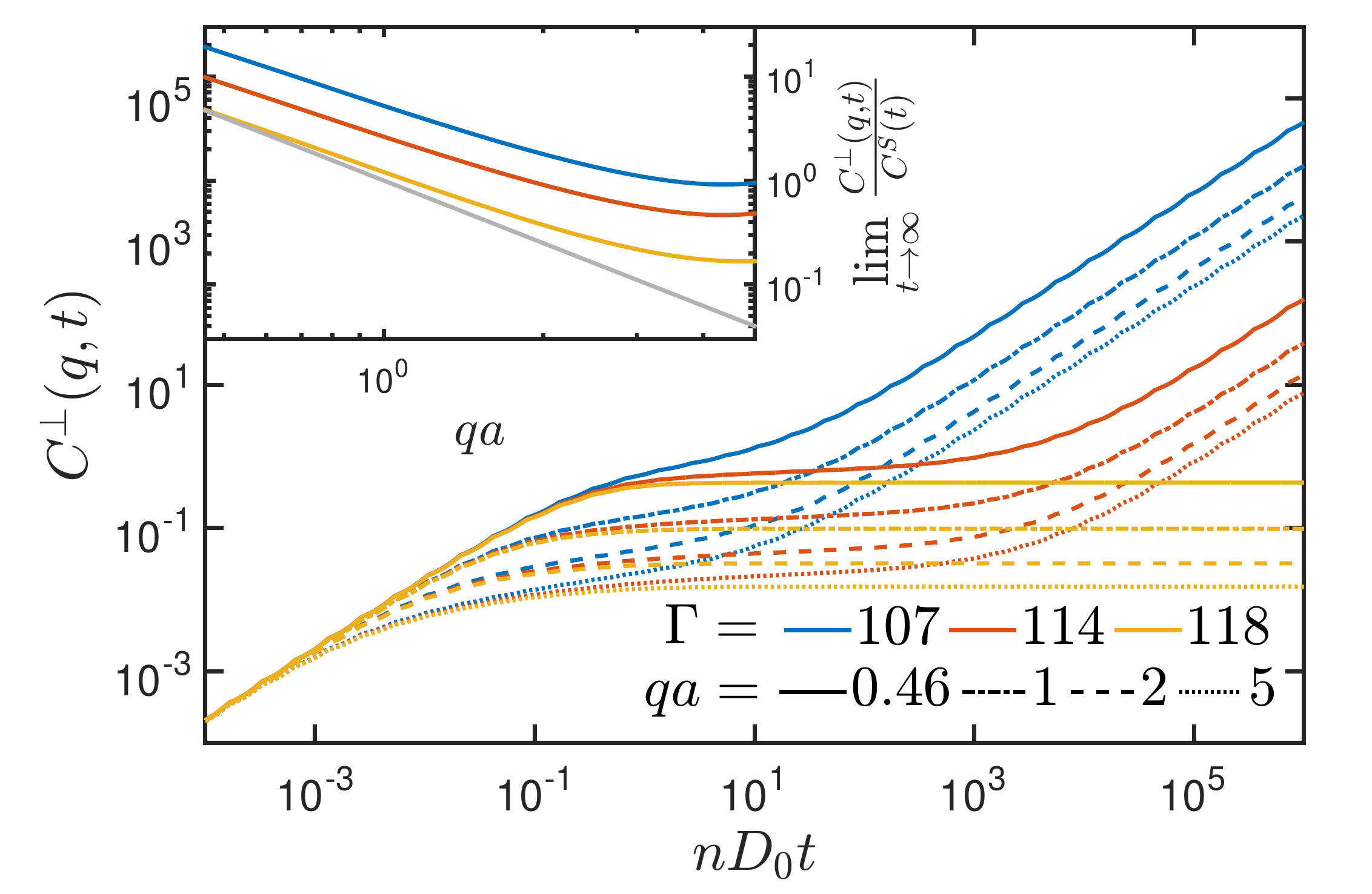}
\caption{ Transversal collective mean-squared displacements  $ C^\perp(q,t)$ from MCT for various wavevectors $q$ as labeled and for three different values of $\Gamma$. While $\Gamma=107$ and $114$ are fluid states, $\Gamma=118$ is a glass in this one-component dipolar system. The inset shows that for wavevectors $qa\ll1$ the $1/q^2$-behavior (grey solid line) predicted by the generalized hydrodynamics result $C^\perp_{gH}(q,t)\to C^s(t) / q^2$ from Eq.~\eqref{e5} is approached at long times for both glass and fluid. Since the three curves collapse for $qa\ll1$ the fluid curves are shifted by factor 2 and 4 respectively for visibility.}
\label{fig2}
\end{figure}

The overdamped equations of motion of the (complete tensorial) CMSD were given in Ref.~\cite{Klix2012}. They rest on {\it i)} the link between displacement and velocity $\dot{\bf u}_i={\bf v}_i$ of particle $i$ presumed valid in fluid and solid states \footnote{In crystals, the time derivative of the displacement field contains the velocity but also a dissipative (isothermal) coupling to the deviatoric stress field: $\partial_t {\bf u}(\rb,t)= {\bf v}(\rb,t) - {\bf \zeta} \nabla ( \boldsymbol{ \sigma}(\rb,t)-p{\bf 1})$  \cite{Martin1972}; see \cite{Szamel2011} concerning  displacements as excitations from quasi-equilibrium positions.} and on {\it ii)} results for velocity correlations obtained by liquid theory  \cite{Hansen}.
Figure~\ref{fig2} shows typical curves for the TCMSD numerically obtained employing approximations familiar from MCT in order to evaluate the arising memory kernels \cite{SI}. The calculation mimics a two-dimensional one-component system of dipolar Brownian particles, which undergoes a glass transition at $\Gamma^{\rm MCT}_g=115$ \cite{Hajnal2011} \footnote{The difference of $\Gamma_g$ between MCT and experiment rests partly on the number of components, but also on the known aspect of MCT to overestimate the tendency to vitrification \cite{goetze, Hajnal2011}.}.  The curves exhibit a scaling limit of generalized hydrodynamics  for small wavevectors, which is of central interest in order to obtain the far-field strain behavior.

The generalized hydrodynamics appropriate in supercooled fluids is obtained from taking the limit of long wavelength fluctuations $qa \ll 1$ and keeping the possibility for slow dynamics \cite{Latz}. Our result, obtained within the Zwanzig-Mori projection operator formalism, describes initially diffusive particle displacements growing linearly in time (with $D_0$ the dilute diffusion coefficient and friction coefficient $\zeta_0=k_BT/D_0$). With increasing time the diffusive displacements get hindered by  interactions captured in a retarded friction kernel:
\begin{equation}\label{e4}
 C^\perp_{gH}(q,t)+ \frac{q^2}{\zeta_0 n}\; \int_0^tdt'\;  G^{\perp}(t-t')  C^\perp_{gH}(q,t') = 2D_0\, t\; ,
\end{equation}
where the subscript $_{gH}$ stands for generalized hydrodynamics. The memory kernel contains the potential shear-stresses:  $G^\perp(t)= \frac{n}{k_BT}\; \langle \sigma_\perp(t_{\cal Q})^\ast \sigma_\perp \rangle$.  Its prefactor $q^2$ results from Newton's 2nd law, that forces are transmitted among the particles and sum up to zero in total. The $G^\perp(t)$ is familiar from the theory of transversal momentum fluctuations in liquids, and its integral gives the shear viscosity according to the Green-Kubo relation \cite{Hansen}: $\eta  = \int_0^\infty dt\,  { G}^{\perp}(t) $. It differs from the stress auto-correlation function in a solid \cite{Wittmer2015}.

The equation of motion for the transversal collective mean-squared displacements contains  a length $L$ defined via:  $(qL)^2= \frac{q^2 \eta}{n \zeta_0}$,  which determines the behavior at long times \footnote{At long times, a Markovian approximation to the memory kernel is valid, which explains the occurrence of the shear viscosity in $L$.}. In the normal hydrodynamic description of a liquid, the limit of small wavevector is taken which leads to $qL\ll1$.Then displacements grow diffusively for all times, $C^{\perp}_{gH}(q\ll1/L,t)= 2 D_0 t$, and the strain correlation function $C_{xy}(\rb,t)$ decays on local distances. Not surprisingly, an equilibrium liquid does not support long ranged strain correlations as present in elastic systems.  The crossover to simple hydrodynamics happens at $q < 1/L$, with $L$  of the order of the average particle distance $a$ in low-viscous fluids \footnote{Note, that this crossover is outside the wavevector grid used in the MCT numerical
calculations of Fig.~\ref{fig2}.}: $L/a=\sqrt{{\eta}/({\zeta_0 n a^2})}$ (which is $\sqrt{\eta/\zeta_0}$ in two-dimensions). This wavevector also limits the range where transversal sound waves are seen in supercooled molecular systems \cite{Ahluwalia1998,Torchinsky2012}. In a glass, strain fluctuations are tested at times $t$ shorter than $\tau$ before the stresses have relaxed. Then the shear stress memory kernel takes the value of the shear modulus $\mu$ \cite{Klix2012}, and the variance of displacements increases with wavevector like a power-law: $C^\perp_{gH}(q, t\ll \tau)= \frac{2 k_BT n}{\mu q^2}$. The small-$q$ divergence of the variance of displacements is the origin of the far-field power-law in Eshelby's result.  Equation~\eqref{e2} holds with $C^s(t)=2 k_BT n/\mu$.

\begin{figure}[tb]
\includegraphics[width=\linewidth]{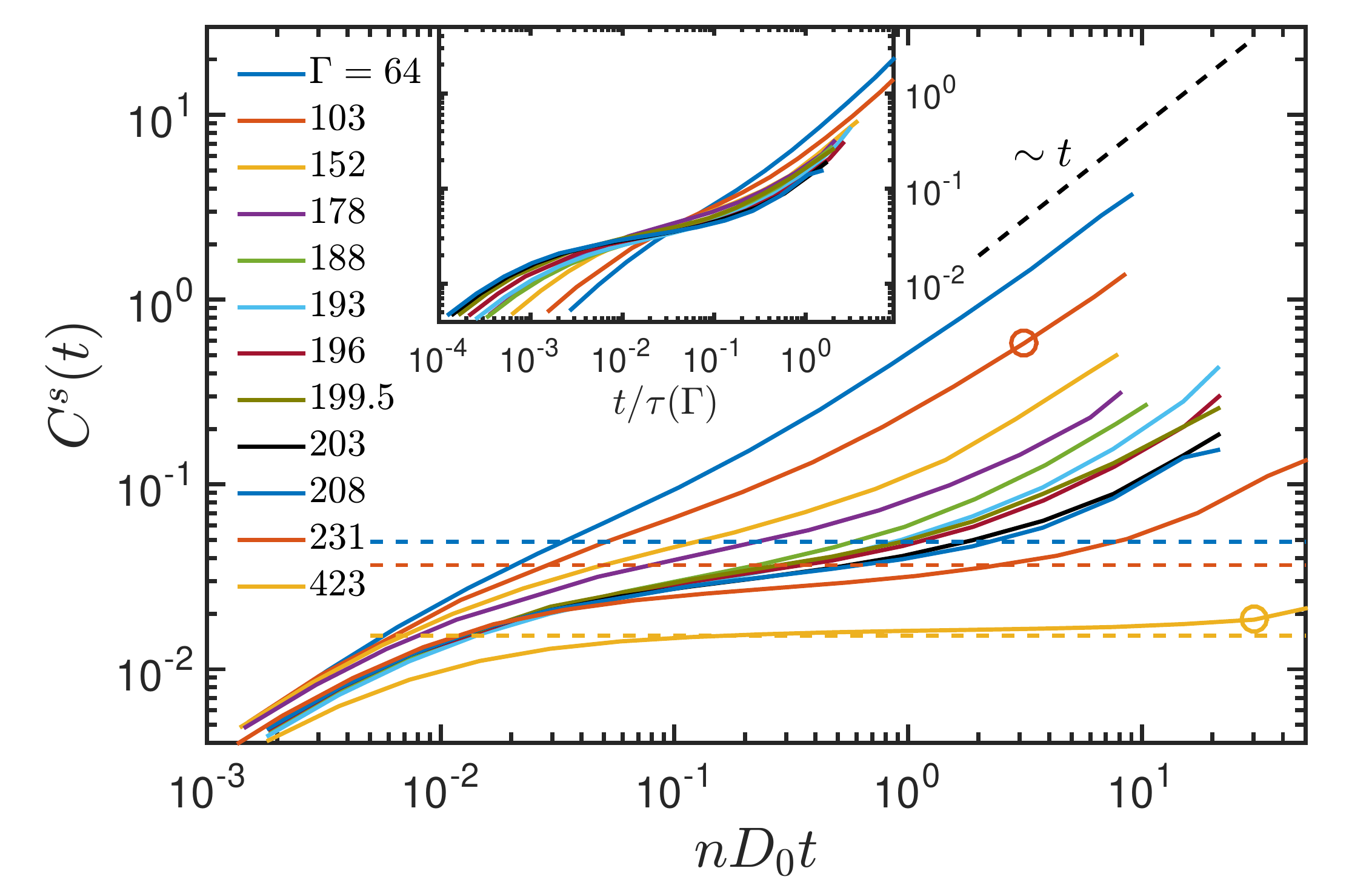}
\includegraphics[width=\linewidth]{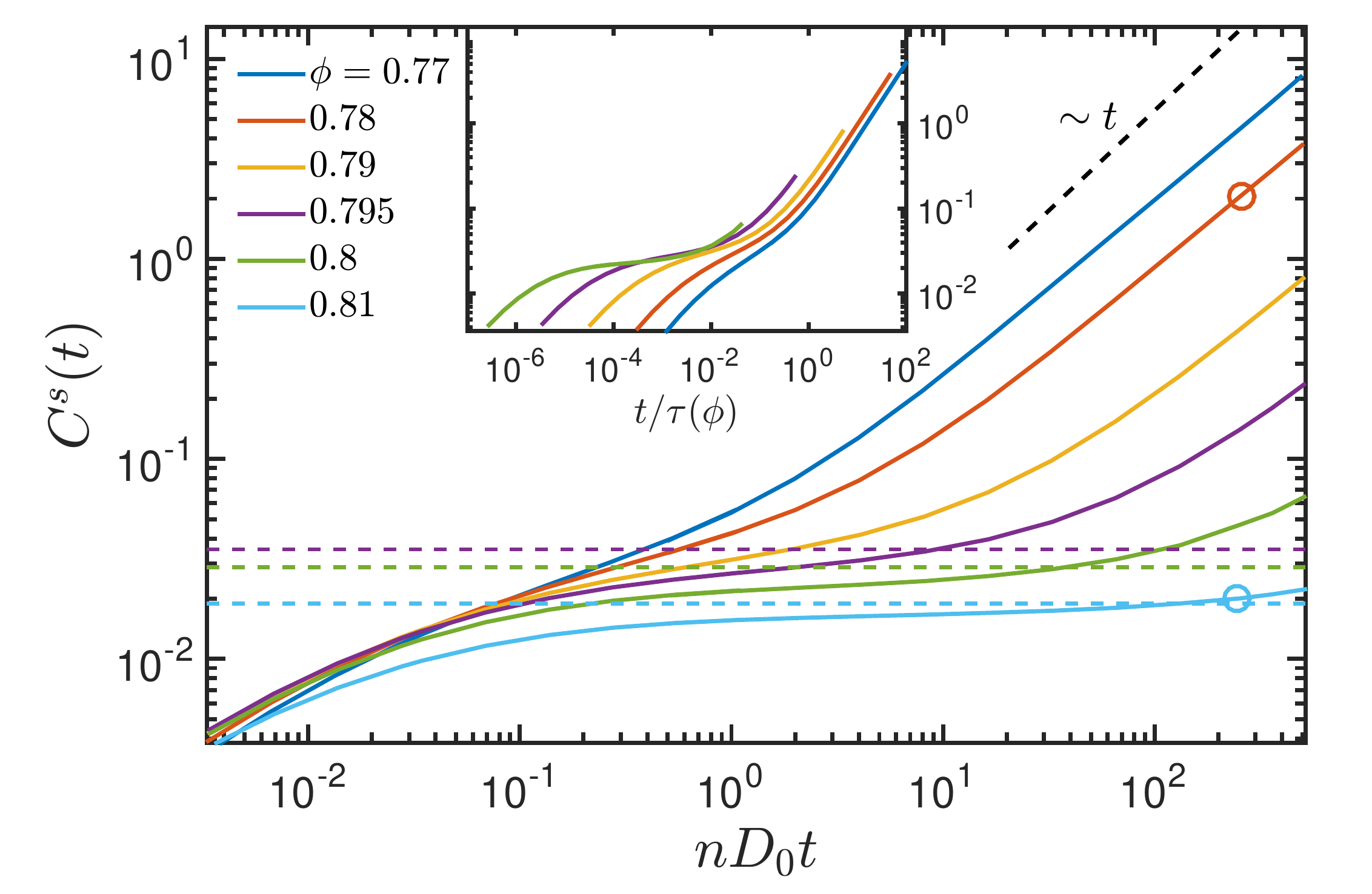}
\caption{Amplitude function $C^s(t)$ of the far-field power-law decay of transversal mean-squared strain fluctuations given in Eq.~\eqref{e2}. It gives the strength of the algebraic $1/r^2$ power-law decay with $\cos{(4\theta)}$- symmetry in $C_{xy}(\rb,t)$, which holds for $a\ll r\ll L$ according to Eq.~\eqref{e5}. The upper panel gives data measured in the colloidal layers, the lower panel gives corresponding data measured by BD simulations in a binary mixture of hard disks \cite{SI}. Legends give the inverse temperatures $\Gamma$ or the packing fractions $\phi$ spanning from fluid to glass states. Circles mark the times, where Eshelby-strain patterns are shown in the inset of Fig.~\ref{fig1} (and Fig.~5 in the SI \cite{SI}). Dashed lines give the elastic limits $C^s(t\to\infty)=2k_BT n({1}/{\mu}-{1}/{\mu^\|})$ with moduli obtained from the dispersion relations following Ref.~\cite{Klix2012}. The insets show asymptotic collapse of the fluid curves when plotted versus rescaled time $t/\tau$ with the final relaxation times $\tau$ obtained from density correlation functions \cite{SI}.}
\label{fig3}
\end{figure}

The increase of the viscosity opens an additional spatio-temporal window, where the shear modulus has decayed to zero, yet the displacements still diverge with $q^{-2}$. This `supercooled liquid' regime is characterized by $t\gg \tau$ and $qL\gg 1$, which requires high values of the viscosity. Then the TCMSD obeys:  $C^{\perp}_{gH}(q\gg 1/L,t\gg\tau)= \frac{2 k_BT n}{\eta q^2} t$, and   the strain correlation function exhibits the spatial behavior reminiscent of solid-like behavior,  $C_{xy}(\rb,t) \to \cos{4\theta} \frac{2k_BT t}{\eta r^2}$. Yet, the coefficient of the spatial power-law is viscous and not elastic. The long-ranged correlation of strains arises from momentum-conservation  during the particle interactions, which causes the slow collective transport  of momentum in the viscoelastic particle system to dominate over the local friction intrinsic in the Langevin description:  $(qL)^2= \frac{q^2 \eta/n}{\zeta_0 } \gg 1$.

The glass and supercooled regimes of strain fluctuations can be summarized in a  scaling law  \footnote{MCT actually predicts also a second $\beta$-scaling law beyond the $\alpha$-scaling law of Eq.~\eqref{e5}.}. The coefficient $C^s(t)$ of the far-field strain-decay in Eq.~\eqref{e2} can
be obtained from a limiting solution of Eq.~\eqref{e4} valid for $qL\gg1$ and $tD_0n\gg1$ in incompressible systems:
\begin{equation}\label{e5}
C^\perp_{gH}(q,t) \to  \frac{C^s(t)}{q^2} \; \mbox{with}\; C^s(t) \to \left\{ \begin{array}{ll}
\frac{2 k_BT n}{\mu}  & t \ll \tau \\
\frac{2 k_BT n t}{\eta} & t\gg \tau
\end{array}\right.
\end{equation}
The equation for $C^s(t)$ follows from Eq.~\eqref{e4} by neglecting the first term. The ansatz of an exponentially decaying memory kernel $G^\perp=\mu e^{-t/\tau}$ would  capture both asymptotes on the right hand side of Eq.~\eqref{e5}: In analogy to a Maxwell fluid under load, $\mu$ would appear for short times and $\eta/t$ for long ones.

Figure \ref{fig3} contains the measured $C^s(t)$ obtained from the real-space analysis considering the far-field power-law decay of the mean-squared strains shown in Fig.~\ref{fig1} \cite{SI}. Additional data sets around the glass transition (inverse) temperature $\Gamma_g\approx 200$ are added.  In the lower panel $C^s(t)$ data obtained identically from Brownian dynamics (BD) simulations of binary hard disks
are shown. The  system's glass transition packing fraction lies at $\phi_g\approx 0.795$
\cite{Weysser2011}. The measured data from experiment and simulation qualitatively agree and exhibit the predicted behaviors. Over a time window increasing when approaching the glass transition, the far-field amplitude   $C^s(t)$ arrests on a constant value as corresponds to elastic solid-like behavior. The elastic moduli determined independently, closely match the plateau values during the intermediate time-window. In fluid states, the far-field amplitude increases for long times asymptotically linearly in time with prefactor given by the inverse viscosity, or equivalently the final relaxation time $\tau$. Rescaling the $C^s(t)$ during this final process using $\tau$ obtained a priori from the density correlation functions (see SI \cite{SI}) gives a satisfactory collapse of the curves in fluid states; see the insets of Fig.~\ref{fig3}.
The BD data would collapse far better if  replotted versus $nD_Lt$ (not shown), where $D_L$ is the long-time self-diffusion coefficient.  This may indicate that strain fluctuations decouple from the structural relaxation like familiar for diffusion, which is often taken as indication for heterogeneous dynamics \cite{Rizzo2015}.

In summary, we have shown that hexadecupolar Eshelby-strain-correlations which are reminiscent of standard elastic behavior can also be detected in experiment for a supercooled fluid when the shear elasticity has already decayed to zero in the long time limit. An analysis of the retarded  mean-squared strain  fluctuation functions within the framework of Mode-Coupling-Theory can explain this: For sufficiently large viscosities an additional spatio-temporal window opens where correlated displacements diverge with lengthscales squared ($~ q^{-2}$) while elastic correlations have already decayed. The origin of the long-ranged strain fluctuations is the conservation of momentum in particle interactions. This derivation within generalized hydrodynamics of supercooled liquids opens the  perspective to connect to the theories of plastic flow in low-temperature glasses. E.g.~the connection of our viscoelastic scaling law to  inherent structures  remains to be established and promises insights into the dynamics on
the potential energy landscape at high temperatures \cite{Chowdhury2016}. Our theoretical results in Fourier-space hold in two and three dimensions, and rationalize data from colloidal layers, even though computer simulations of single particle motion indicated that localization in two dimensional glasses is rather weak \cite{Flenner2015}. Accordingly, our simulations require large systems in order to observe the far-field behavior.

\begin{acknowledgments}
We thank J.-L. Barrat for helpful discussions.
P.K. acknowledges financial support from the Young Scholar Fund, University of Konstanz.
\end{acknowledgments}
\bibliographystyle{apsrev4-1}
\bibliography{lit}
\appendix

\end{document}